\documentclass[%
 reprint,
superscriptaddress,
 amsmath,amssymb,
pra,https://it.overleaf.com/project/5f1efe58c74a680001ebd365
]{revtex4-2}

\usepackage[utf8]{inputenc}
\usepackage[T1]{fontenc}
\usepackage{graphicx}
\usepackage{dcolumn}
\usepackage{bm}
\usepackage{tablefootnote}
\usepackage{xcolor}
\usepackage{fourier}
\usepackage{orcidlink}
\usepackage{float}
\usepackage{threeparttable}

\usepackage{placeins}
\usepackage{color}

\usepackage{xcolor}
\usepackage{capt-of}
\usepackage{nicefrac}

\begin{document}

\author{F.P. Gustafsson\orcidlink{0000-0003-2063-4310}}
\email{fredrik.parnefjord.gustafsson@cern.ch}
\affiliation{Instituut voor Kern- en Stralingsfysica, KU Leuven, 3001 Leuven, Belgium}
\affiliation{CERN, 1211 Geneva 23, Switzerland}

\author{L. V. Rodr\'iguez\orcidlink{0000-0002-0093-2110}}
\email{liss.vazquez.rodriguez@cern.ch}
\affiliation{Max-Planck-Institut f\"ur Kernphysik, Saupfercheckweg 1, 69117 Heidelberg, Germany}
\affiliation{CERN, 1211 Geneva 23, Switzerland}
\affiliation{IJCLab, Université Paris-Saclay, CNRS/IN2P3, 91405 Orsay, France}

\author{R.F. Garcia Ruiz\orcidlink{0000-0002-0093-2110}}
\email{rgarciar@mit.edu}
\affiliation{Massachusetts Institute of Technology, Cambridge, MA 02139, USA}
\affiliation{School of Physics and Astronomy, The University of Manchester, Manchester, M13 9PL, United Kingdom}
\affiliation{CERN, 1211 Geneva 23, Switzerland}

\author{T. Miyagi\orcidlink{0000-0002-6529-4164}}
\affiliation{Center for Computational Sciences, University of Tsukuba, 1-1-1 Tennodai, Tsukuba, Ibaraki 305-8577, Japan}
\affiliation{Institut f\"ur Kernphysik, Technische Universit\"at Darmstadt, 64289 Darmstadt, Germany}
\affiliation{ExtreMe Matter Institute, GSI Helmholtzzentrum f\"ur Schwerionenforschung GmbH, 64291 Darmstadt, Germany}
\affiliation{Max-Planck-Institut f\"ur Kernphysik, Saupfercheckweg 1, 69117 Heidelberg, Germany}

\author{S.W. Bai}
\affiliation{School of Physics and State Key Laboratory of Nuclear Physics and Technology, Peking University, Beijing 100871, China}

\author{D.L.~Balabanski\orcidlink{0000-0003-2940-014X}}
\affiliation{ELI-NP, Horia Hulubei National Institute for R\&D in Physics and Nuclear Engineering, 077125 Magurele, Romania}

\author{C.L. Binnersley}
\affiliation{School of Physics and Astronomy, The University of Manchester, Manchester, M13 9PL, United Kingdom}

\author{M.L. Bissell}
\affiliation{School of Physics and Astronomy, The University of Manchester, Manchester, M13 9PL, United Kingdom}

\author{K. Blaum\orcidlink{0000-0003-4468-9316}}
\affiliation{Max-Planck-Institut f\"ur Kernphysik, Saupfercheckweg 1, 69117 Heidelberg, Germany} 

\author{B. Cheal\orcidlink{0000-0002-1490-6263}}
\affiliation{Oliver Lodge Laboratory, Oxford Street, University of Liverpool, Liverpool L69 7ZE, United Kingdom}

\author{T.E. Cocolios\orcidlink{0000-0002-0456-7878}}
\affiliation{Instituut voor Kern- en Stralingsfysica, KU Leuven, 3001 Leuven, Belgium}

\author{G.J. Farooq-Smith\orcidlink{0000-0001-8384-7626}}%
\affiliation{Instituut voor Kern- en Stralingsfysica, KU Leuven, 3001 Leuven, Belgium}

\author{K.T. Flanagan}
\affiliation{School of Physics and Astronomy, The University of Manchester, Manchester, M13 9PL, United Kingdom}
\affiliation{Photon Science Institute, Alan Turing Building, The University of Manchester, Manchester, M13 9PY, United Kingdom}

\author{S. Franchoo\orcidlink{0000-0001-7520-5922}}
\affiliation{IJCLab, Université Paris-Saclay, CNRS/IN2P3, 91405 Orsay, France}

\author{A. Galindo-Uribarri\orcidlink{0000-0001-7450-404X}}
\affiliation{Physics Division, Oak Ridge National Laboratory, Oak Ridge, Tennessee, 37831, USA}
\affiliation{Department of Physics and Astronomy, University of Tennessee, Knoxville, Tennessee 37916, USA}

\author{G. Georgiev\orcidlink{0000-0003-1467-1764}}
\affiliation{IJCLab, Université Paris-Saclay, CNRS/IN2P3, 91405 Orsay, France}

\author{W. Gins\orcidlink{0000-0002-2353-7455}}
\affiliation{Instituut voor Kern- en Stralingsfysica, KU Leuven, 3001 Leuven, Belgium}

\author{C. Gorges}
\affiliation{Institut f\"ur Kernphysik, Technische Universit\"at Darmstadt, 64289 Darmstadt, Germany}

\author{R.P. de Groote\orcidlink{0000-0003-4942-1220}}
\affiliation{Instituut voor Kern- en Stralingsfysica, KU Leuven, 3001 Leuven, Belgium}

\author{H.~Heylen\orcidlink{0009-0001-7278-2115}}
\affiliation{CERN, 1211 Geneva 23, Switzerland}

\author{J.D. Holt\orcidlink{0000-0003-4833-7959}}
\affiliation{TRIUMF, Vancouver, British Columbia V6T 2A3, Canada}
\affiliation{
  Department of Physics, McGill University, Montr\'eal, QC H3A 2T8, Canada
  }

\author{A. Kanellakopoulos\orcidlink{0000-0002-6096-6304}}
\affiliation{Instituut voor Kern- en Stralingsfysica, KU Leuven, 3001 Leuven, Belgium}

\author{J. Karthein\orcidlink{0000-0002-4306-9708}}
\thanks{Present Address: Texas A\&M University, Cyclotron Institute / Dep. of Physics \& Astronomy, College Station, TX 77840, USA}
\affiliation{Massachusetts Institute of Technology, Cambridge, MA 02139, USA}

\author{S. Kaufmann}
\affiliation{Institut f\"ur Kernphysik, Technische Universit\"at Darmstadt, 64289 Darmstadt, Germany}

\author{\'A. Koszor\'us\orcidlink{0000-0001-7959-8786}}
\affiliation{Instituut voor Kern- en Stralingsfysica, KU Leuven, 3001 Leuven, Belgium}

\author{K. K\"onig\orcidlink{0000-0001-9415-3208}}%
\affiliation{Institut f\"ur Kernphysik, Technische Universit\"at Darmstadt, 64289 Darmstadt, Germany}

\author{V.~Lagaki}
\affiliation{CERN, 1211 Geneva 23, Switzerland}
\affiliation{Institut für Physik, Universit\"at Greifswald, 17487 Greifswald, Germany}

\author{S. Lechner\orcidlink{0000-0002-1695-5683}}
\affiliation{CERN, 1211 Geneva 23, Switzerland}
\affiliation{Technische Universit\"at Wien, Karlsplatz 13, 1040 Wien, Austria}

\author{B. Maass\orcidlink{0000-0002-6844-5706}}
\affiliation{Institut f\"ur Kernphysik, Technische Universit\"at Darmstadt, 64289 Darmstadt, Germany}

\author{S. Malbrunot-Ettenauer}
\affiliation{CERN, 1211 Geneva 23, Switzerland}

\author{W. Nazarewicz\orcidlink{0000-0002-8084-7425}}
\affiliation{Facility for Rare Isotope Beams \& Department of Physics and Astronomy, Michigan State University, East Lansing, MI, USA}

\author{R. Neugart}
\affiliation{Institut für Kernchemie, Universit\"at Mainz, 55128 Mainz, Germany}
\affiliation{Max-Planck-Institut f\"ur Kernphysik, Saupfercheckweg 1, 69117 Heidelberg, Germany}

\author{G.~Neyens\orcidlink{0000-0001-8613-1455}}
\affiliation{Instituut voor Kern- en Stralingsfysica, KU Leuven, 3001 Leuven, Belgium}
\affiliation{CERN, 1211 Geneva 23, Switzerland}

\author{W. Nörtershäuser\orcidlink{0000-0001-7432-3687}}
\affiliation{Institut f\"ur Kernphysik, Technische Universit\"at Darmstadt, 64289 Darmstadt, Germany}

\author{T. Otsuka\orcidlink{0000-0002-1593-5322}}
\affiliation{Center for Nuclear Study, University of Tokyo, Hongo, Bunkyo-ku, Tokyo 113-0033, Japan}
\affiliation{Instituut voor Kern- en Stralingsfysica, KU Leuven, 3001 Leuven, Belgium}
\affiliation{Department of Physics, University of Tokyo, Hongo, Bunkyo-ku, Tokyo 113-0033, Japan}
\affiliation{RIKEN Nishina Center, 2-1 Hirosawa, Wako, Saitama 351-0198, Japan}
\affiliation{National Superconducting Cyclotron Laboratory, Michigan State University, East Lansing, Michigan 48824, USA}

\author{P.-G. Reinhard\orcidlink{0000-0002-4505-1552}}
\affiliation{Institut f\"ur Theoretische Physik, Friedrich-Alexander-Universit\"at, Erlangen/N\"urnberg, Germany}

\author{N. Rondelez\orcidlink{0000-0001-8356-9988}}
\affiliation{Instituut voor Kern- en Stralingsfysica, KU Leuven, 3001 Leuven, Belgium}

\author{E. Romero-Romero\orcidlink{0000-0003-4015-4904}}%
\affiliation{Physics Division, Oak Ridge National Laboratory, Oak Ridge, Tennessee, 37831, USA}
\affiliation{Department of Physics and Astronomy, University of Tennessee, Knoxville, Tennessee 37916, USA}

\author{C.M. Ricketts}
\affiliation{School of Physics and Astronomy, The University of Manchester, Manchester, M13 9PL, United Kingdom}

\author{S. Sailer}
\affiliation{Technische Universit\"at München, 80333 Munich, Germany}

\author{R. S\'anchez\orcidlink{0000-0002-4892-4056}}
\affiliation{GSI Helmholtzzentrum f\"ur Schwerionenforschung GmbH, 64291 Darmstadt, Germany}

\author{S. Schmidt}
\affiliation{Institut f\"ur Kernphysik, Technische Universit\"at Darmstadt, 64289 Darmstadt, Germany}

\author{A. Schwenk\orcidlink{0000-0001-8027-4076}}
\affiliation{Institut f\"ur Kernphysik, Technische Universit\"at Darmstadt, 64289 Darmstadt, Germany}
\affiliation{ExtreMe Matter Institute, GSI Helmholtzzentrum f\"ur Schwerionenforschung GmbH, 64291 Darmstadt, Germany}
\affiliation{Max-Planck-Institut f\"ur Kernphysik, Saupfercheckweg 1, 69117 Heidelberg, Germany}

\author{S.R. Stroberg}
\affiliation{TRIUMF, Vancouver, British Columbia V6T 2A3, Canada}
\affiliation{Department of Physics, University of Washington, Seattle WA, USA}
\affiliation{Physics Division, Argonne National Laboratory, Lemont IL, USA}

\author{N. Shimizu}
\affiliation{Center for Computational Sciences, University of Tsukuba,
1-1-1 Tennodai, Tsukuba, Ibaraki 305-8577, Japan}

\author{Y. Tsunoda\orcidlink{0000-0002-0398-1639}}
\affiliation{Center for Nuclear Study, University of Tokyo, Hongo, Bunkyo-ku, Tokyo 113-0033, Japan}

\author{A.R.~Vernon}
\affiliation{Massachusetts Institute of Technology, Cambridge, MA 02139, USA}

\author{L. Wehner}
\affiliation{Institut für Kernchemie, Universit\"at Mainz, 55128 Mainz, Germany}

\author{S. G. Wilkins}
\affiliation{Massachusetts Institute of Technology, Cambridge, MA 02139, USA}

\author{C. Wraith}
\affiliation{Oliver Lodge Laboratory, Oxford Street, University of Liverpool, Liverpool L69 7ZE, United Kingdom}

\author{L. Xie}
\affiliation{School of Physics and Astronomy, The University of Manchester, Manchester, M13 9PL, United Kingdom}

\author{Z. Y. Xu}
\affiliation{Instituut voor Kern- en Stralingsfysica, KU Leuven, 3001 Leuven, Belgium}

\author{X.F. Yang\orcidlink{0000-0002-1633-4000}}
\affiliation{School of Physics and State Key Laboratory of Nuclear Physics and Technology, Peking University, Beijing 100871, China}

\author{D. T. Yordanov\orcidlink{0000-0002-1592-7779}}
\affiliation{Max-Planck-Institut f\"ur Kernphysik, Saupfercheckweg 1, 69117 Heidelberg, Germany}
\affiliation{IJCLab, Université Paris-Saclay, CNRS/IN2P3, 91405 Orsay, France}
\affiliation{CERN, 1211 Geneva 23, Switzerland}

\title{Charge Radii Measurements of Exotic Tin Isotopes in the Proximity of $N=50$ and $N=82$}

\begin{abstract}
We report nuclear charge radii for the isotopes $^{104-134}$Sn, measured using two different collinear laser spectroscopy techniques at ISOLDE-CERN. These measurements clarify the arch-like trend in charge radii along the isotopic chain and reveal an odd-even staggering that is more pronounced near the $N=50$ and $N=82$ shell closures. The observed local trends are well described by both nuclear density functional theory and valence space in-medium similarity renormalization group calculations. Both theories predict appreciable contributions from beyond-mean-field correlations to the charge radii of the neutron-deficient tin isotopes. The models, however, fall short of reproducing the magnitude of the known $B(E2)$ transition probabilities, highlighting the remaining challenges in achieving a unified description of both ground-state properties and collective phenomena.\pagebreak
\end{abstract}

\maketitle

\textit{Introduction}.-- The mean-square charge radius is a fundamental property of the atomic nucleus, providing direct insight into its size and the spatial distribution of nuclear electric charge. Precise measurements of its evolution along isotopic chains offer unique insight into  nuclear shell structure, collective phenomena, and the underlying forces. Understanding how single-nucleon and collective motion emerges from microscopic interactions between protons and neutrons is an essential step towards a unified description of the atomic nucleus. Collective nuclear properties are commonly enhanced in open-shell nuclei and significantly reduced around closed nuclear shells. In this context, the tin (Sn) isotopic chain, with its closed proton shell at $Z=50$, is of particular importance. Currently, it offers a long sequence of isotopes available for study, stretching across the major neutron shell closures at $N=50$ and $N=82$. This unique landscape makes the tin chain a fertile testing ground for the development of theoretical ab initio models~\cite{hin12,Morris2018,togashi2018,gys19,mou21,Tichai:2023epe,Arthuis:2024mnl}. Many properties of nuclei in the vicinity of the doubly magic nuclei are expected to be accurately predicted by considering only a few valence nucleons. While the doubly magic nature of $^{132}$Sn has been well-established experimentally~\cite{jones2010magic,Rosiak2018a,Gorges2019,rodriguez2020doubly,Ver22}, significantly less data are available to understand the evolution of nuclear structure properties in the region from mid-shell down towards $^{100}$Sn. This is due to the considerable experimental challenges associated with producing and studying these neutron-deficient isotopes. 
The study of these isotopes is crucial as contradictory experimental evidence has been reported about the evolution of single-particle properties approaching $^{100}$Sn~\cite{Se07,Dar10,Ba05}, and several questions remain in connection with the robustness of the $N = Z = 50$ shell closures~\cite{hin12,Ba05,PhysRevLett.98.172501,Va07,reponen2021evidence,mou21,Nies2023,Ge2024}. While the largest Gamow-Teller strength among all allowed $\beta$-decays was found in $^{100}$Sn and supports a dominant single-particle behavior~\cite{hin12}, the existence of low-lying energy levels and relatively large $B(E2)$ values, known down to $^{104}$Sn~\cite{PhysRevLett.98.172501,Va07,back2013transition,guastalla2013,doornenbal2014}, are indicative of a collective picture, in contrast with recent laser spectroscopy results on $_{49}$In~\cite{Karthein2024} and $_{47}$Ag~\cite{reponen2021evidence} isotopes.  

In this work, we present nuclear charge radii measurements for the tin isotopes $^{104-134}$Sn, using two collinear laser spectroscopy (CLS) techniques~\cite{Yan22}. The neutron-deficient isotopes, down to $^{104}$Sn, were investigated with the Collinear Resonance Ionization Spectroscopy (CRIS) setup~\cite{Kos20}, while the neutron-rich isotopes, up to $^{134}$Sn, were measured using the Collinear Laser Spectroscopy (COLLAPS) experiment~\cite{Gorges2019}. Our measurement campaigns thus extend the available data~\cite{blanc2005,eberz1987,Anselment1986} by four isotopes on the neutron-deficient side $^{104-107}$Sn, while completing the earlier reported radii in~\cite{Gorges2019} with the odd-even isotopes including $^{133}$Sn \footnote{We note that Fig.\,7 of  Ref.~\cite{Reinhard.2024} contains a data point for the charge radius of $^{133}$Sn. This has been a tentative estimate, not based on actual measurement, and was unfortunately not removed from the final version of the paper.}. This exceptionally long dataset, spanning thirty-one isotopes, serves to reveal intricate changes in the mean-square charge radii towards the neutron shell closures, which can be explored by nuclear density functional theory (DFT) ~\cite{Bender2003,nucleardft} and ab initio valence space in-medium similarity renormalization group (VS-IMSRG) calculations~\cite{Hergert:2015awm,Stro19ARNPS}. 

\textit{Experiment}.-- Two CLS methods were employed to measure the charge radii of tin isotopes. Neutron-deficient isotopes, from $^{124}$Sn down to $^{104}$Sn, were measured at CRIS using two recently developed ionization schemes for probing the two atomic transitions $5s^{2}5p^{2} \; ^{3\!}P_{1} \rightarrow 5s^{2}5p6s\;^{3\!}P_{2}$ (284\,nm) and $5s^{2}5p^{2} \; ^{1\!}S_{0} \rightarrow 5s^{2}5p7s \; ^{1\!}P_{1} $ (281\,nm) \cite{Fredrik2020}. Neutron-rich isotopes, from $^{108}$Sn up to $^{134}$Sn were investigated at COLLAPS using two complementary transitions, $5s^{2}5p^{2} \; ^{1\!}S_{0} \rightarrow 5p6s \; ^{1\!}P_{1}$ (452.5\,nm) and $5s^{2}5p^{2} \; ^{3\!}P_{0} \rightarrow 5p6s \; ^{3\!}P_{1} $ (286.3\,nm). These transitions were chosen to maximize sensitivity to nuclear observables. Detailed information about the ion beam production and experimental technique used in the CRIS experiment, as well as the deduced isotope shifts and mean-square charge radii, can be found in the Supplemental Material. For the COLLAPS measurements, all details are described in~\cite{Gorges2019, Dey20, rodriguez2020doubly}. 
The changes in the mean-square nuclear charge radii were extracted from the measured isotope shifts following the relation
\begin{equation}
\delta \nu^{A',A}_i = \nu_i^A - \nu_i^{A^\prime} =  M_{i}\,\dfrac{m_{A}-m_{A'}}{m_{A}m_{A'}}+F_{i}\,\delta\langle r_\mathrm{c}^{2}\rangle^{A',A}\;,
\label{eq:isotopeShift}
\end{equation}
where $F_i$ and $M_i$ are, respectively, the field-shift factor and
the mass-shift factor of the transition $i$, while $m_{A'}$ and $m_{A}$ are the atomic masses of different isotopes.

\begin{table}[]
\caption{Spins and differences in mean-square charge radii of tin isotopes derived from four measured transitions. Statistical uncertainties are indicated in parentheses and systematic contributions are provided in brackets. Details are presented in the Supplemental Material. We note that these $\delta\left\langle r_\mathrm{c}^{2}\right\rangle$ values \textit{must not be combined} with the ones previously published for the even-even isotopes in Ref.\,\cite{Gorges2019} since those are also included in the present analysis. Isotopes measured for the first time in this work have been indicated with an asterisk.}
\label{table:DeltaR}

\begin{tabular}{cccc}
 &  &  &   \\ \cline{1-4}\hline\hline
$A$ & $I^{\pi}$ & $\delta\langle r_\mathrm{c}^{2}\rangle^{124,A}$ (fm$^{2}$) &  ${\langle r_\mathrm{c}^{2} \rangle}^{1/2}$ (fm)   \\ \cline{1-4} \hline\hline
104* & 0 & $-1.497$ (5) [41] & $4.512$ (5)  \\
105* & 5/2$^+$ & $-1.408$ (4) [38] & $4.522$ (4)   \\
106* & 0 & $-1.257$ (3) [25] & $4.539$ (3) \\
107* & 5/2$^+$ & $-1.180$ (3) [24] & $4.547$ (3)  \\
108 & 0 & $-1.072$ (3) [16] & $4.559$ (2)   \\
109 & 5/2$^+$ & $-1.002$ (1) [16] & $4.567$ (2)   \\
110 & 0 & $-0.901$ (3) [10] & $4.578$ (2) \\
111 & 7/2$^+$ & $-0.848$ (1) [12] & $4.583$ (2) \\
112 & 0 & $-0.747$ (1) [6] & $4.594$ (1)   \\
113 & 1/2$^+$ & $-0.685$ (2) [7] & $4.601$ (1)  \\
114 & 0 & $-0.606$ (1) [9] & $4.610$ (1)   \\
115 & 1/2$^+$ & $-0.560$ (1) [5] & $4.615$ (1)   \\
116 & 0 & $-0.462$ (1) [7] & $4.625$ (1)   \\
117 & 1/2$^+$ & $-0.412$ (1) [4] & $4.631$ (1)  \\
118 & 0 & $-0.324$ (1) [3] & $4.640$ (1)   \\
119 & 1/2$^+$ & $-0.282$ (1) [5] & $4.645$ (1) \\
120 & 0 & $-0.206$ (1) [2] & $4.653$ (1)   \\
121 & 3/2$^+$ & $-0.162$ (1) [5] & $4.658$ (1)   \\
122 & 0 & $-0.097$ (1) [1] & $4.665$ (1)  \\
123 & 11/2$^-$ & $-0.060$ (1) [3] & $4.669$ (1)   \\
124 & 0 & 0 &    \\
125 & 11/2$^-$ & $0.033$ (1) [4] & $4.679$ (1)   \\
126 & 0 & $0.089$ (1) [4] & $4.684$ (1)  \\
127 & 11/2$^-$ & $0.120$ (1) [8] & $4.688$ (1)   \\
128 & 0 & $0.165$ (1) [9] & $4.693$ (1)   \\
129 & 3/2$^+$ & $0.177$ (2) [18] & $4.694$ (2)   \\
130 & 0 & $0.240$ (1) [15] & $4.701$ (2)   \\
131 & 3/2$^+$ & $0.238$ (1) [25] & $4.700$ (3)   \\
132 & 0 & $0.307$ (1) [21] & $4.708$ (2)  \\
133* & 7/2$^-$ & $0.406$ (1) [12] & $4.718$ (2)  \\
134 & 0 & $0.531$ (2) [5] & $4.731$ (1)   \\ \hline\hline
\end{tabular}
\end{table}

The atomic factors, $F_i$ and $M_i$, were extracted from a standard King-plot analysis (as described in Ref.~\cite{Gorges2019}) using muonic data as well as $V^\mathrm{e}_2$ factors derived from elastic electron scattering \cite{50-Sn}. Only the measured isotope shifts of the even stable isotopes were used in the King-plot analysis. Changes in charge radii extracted from the four transitions agree well with each other. The weighted averages were used to determine the nuclear charge radii of the tin isotopes based on the $^{124}$Sn reference value of ${\langle r_\mathrm{c}^{2} \rangle}^{1/2} =4.675\,(1)$  fm of~\cite{50-Sn}. The results are presented in Table~\ref{table:DeltaR}. The uncertainties in the extraction of charge radii are dominated by the atomic factor uncertainty.

\textit{Theory}.-- The measured radii are confronted with predictions utilizing both the VS-IMSRG and DFT methods. The VS-IMSRG results employ two interactions derived from chiral effective field theory, 1.8/2.0 (EM)~\cite{Entem2003,Hebe11fits} and $\Delta$N$^{2}$LO$_{\rm GO}$(394)~\cite{Jia20}. The 1.8/2.0 (EM) interaction was fitted to data of few-nucleon systems $A \leq 4$ and globally reproduces ground-state energies up to $A\sim 100$ nuclei~\cite{Stro21Drip}. This interaction has been shown to provide a reasonably good description of local variations of nuclear charge radii in the medium-mass region~\cite{deGroote2020}, while underestimating their absolute values~\cite{Simo17SatFinNuc}. In contrast, the $\Delta$N$^{2}$LO$_{\rm GO}$(394) interaction includes the $\Delta$-isobar degree of freedom and was fitted to few-body data, selected nuclei, and properties of infinite nuclear matter. This interaction improves the simultaneous reproduction of energies and radii~\cite{Jia20}. Further details of the VS-IMSRG calculations are provided in the Supplemental Material.

The DFT-Fy(IVP) results were obtained with the HFB solver \texttt{SkyAx} for axially symmetric deformed configurations~\cite{SkyAx2021}. The computation of broken-pair states in odd-$N$ isotopes involves the breaking of time-reversal symmetry~\cite{Pototzky2010}. It has been calibrated using the large dataset of ground-state properties of semi-magic nuclei from~\cite{Klupfel2009} augmented by the data on differential charge radii of the calcium~\cite{miller2019a}, tin [$\delta \langle r_\mathrm{c}^2\rangle(^{124}\mathrm{Sn}-^{132}\mathrm{Sn})$] and lead isotopic chains~\cite{Karthein2024}. The calculated charge radii were obtained directly from the charge form factors containing relativistic corrections (including spin-orbit term) and contributions from nucleonic charge form factors \cite{DFTformfactors}. The calibration of the functional produces statistical uncertainties from propagation of the uncertainties of the model parameters~\cite{Klupfel2009,Dobaczewski2014}. Systematic uncertainties come from approximate angular momentum projection and from collective ground-state correlations stemming from low-lying quadrupole excitations~\cite{Kluepfel_2008}. The latter generally increase the radii and so introduce asymmetric uncertainties.

\begin{figure}[h!]
\center
\includegraphics[width=1\columnwidth]{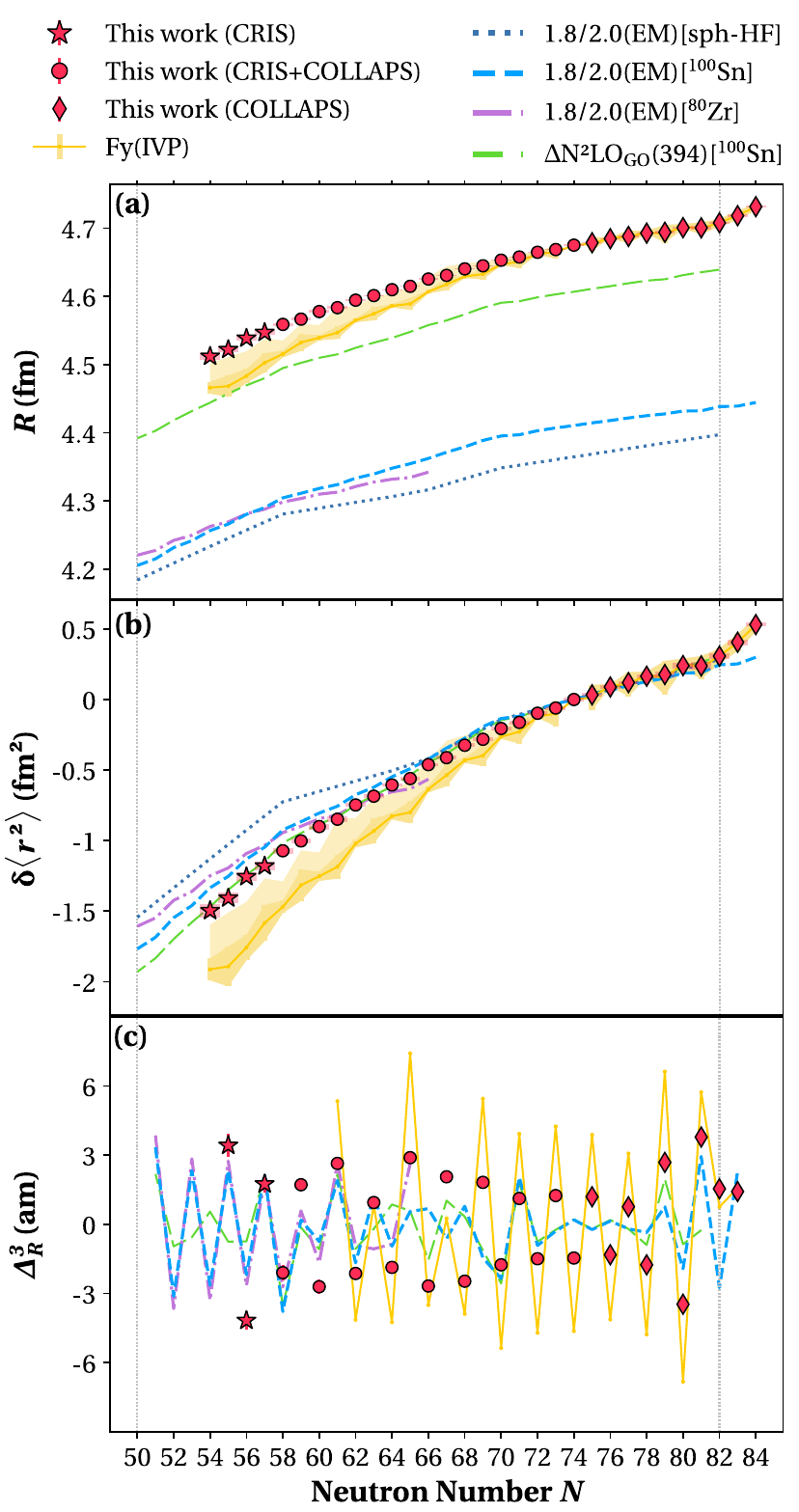}
\caption{ (a) The  absolute root-mean-square radii obtained using $R_c \left({}^{124}\text{Sn}\right) = 4.675\,(1)$~fm \cite{50-Sn}. (b) Differential mean-square charge radii of tin isotopes with respect to $^{124}$Sn. (c) The odd-even staggering $\Delta^{(3)}_{R}=1/2(R_{A+1}+R_{A-1}-2R_{A})$. Experimental results are compared to VS-IMSRG and DFT results  (for details see text). The statistical uncertainties for Fy(IVP) are indicated by dark yellow bands and the systematic uncertainties (mainly ground-state correlations) by light yellow bands.} 
\label{Fig:Radii}
\end{figure}

\textit{Discussion}.-- In Figure~\ref{Fig:Radii}, we present a detailed comparison between our measured charge radii and predictions using both VS-IMSRG and DFT approaches using the experimentally determined spins. 
Figure~\ref{Fig:Radii}(a) reveals that both interactions used in the VS-IMSRG calculations underestimate the absolute charge radii. This underestimation is primarily attributed to an insufficient description of saturation~\cite{Simonis2017,Reinhard2016}. In the description of saturation in VS-IMSRG, three-nucleon (3N) forces play an important role. These cannot be fixed by information from NN scattering but largely cancel in the differential radii $\delta \left\langle r_\mathrm{c}^2 \right\rangle$ that can thus serve as a crucial test for VS-IMSRG calculations. Beyond $N=82$, the VS-IMSRG calculations were performed with a different valence space only for the 1.8/2.0(EM) interaction. As the systematic uncertainties are different in different valence-space calculations, it is challenging to discuss the behavior at $N=82$ in the VS-ISMRG. Indeed, at $N=82$, we observed a $0.12$ fm$^{2}$ difference in the squared radii from the different valence-space calculations.

All modern nuclear DFT models (Skyrme, Fayans, Gogny, relativistic mean field) provide a good description of nuclear radii across the chart of isotopes from $A=16$ on, typically within $\pm 0.02$ fm. They differ in their performance concerning isotopic trends and details in a particular region. Fayans functionals are designed to be more flexible in tuning radius trends than other functionals~\cite{Reinhard2017,Gorges2019}. For other nuclear properties,
their performance is parallel to other well-calibrated DFT models, see Ref.~\cite{ReinhardF2024}.
The Fy(IVP) functional here provides a further improved description of the radii trend and the absolute radii compared to its former variant Fy($\Delta r$,HFB) employed in \cite{Gorges2019} for even-even tin isotopes, particularly in the neutron-rich region and across the kink at the $N=82$ shell-closure.

In Fig.~\ref{Fig:Radii}(b) differential mean-square charge radii relative to $^{124}$Sn are shown. They show an archlike trend towards the known kink at the $^{132}$Sn doubly-magic shell closure. This general behavior superimposed on a linear increase is overall well described by both the DFT and VS-IMSRG calculations. The good agreement with a spherical Hartree-Fock (HF) calculation using the equal-filling approximation (labeled by ``1.8/2.0 (EM) sph-HF'' and discussed below) indicates that this is primarily driven by the bulk radial increase, whereas the correlations included in the VS-IMSRG calculations become important below $N=66$. The collective ground-state correlations become sizable in the DFT calculations in the same region, and they move the result closer to experiment (as indicated with a light yellow band). 

Finally, we study the odd-even staggering $\Delta^{(3)}_{R}=1/2(R_{A+1}+R_{A-1}-2R_{A})$ in  Fig.~\ref{Fig:Radii}(c). The addition of individual neutrons to the nucleus results in appreciable changes to its charge distribution due to pairing and deformation effects~\cite{deGroote2020}. In our data, we observe an increased staggering as we approach the $N=50$ and $N=82$ shell closures. In between these shell closures, we note that the staggering appears more irregular in comparison to that of neutron separation energies, as also seen along the copper chain ~\cite{deGroote2020}, which we attribute to the sensitivity of the charge radius to the occupation of different neutron orbitals with varying spatial extent and polarizability.  In addition, our new measurement of the charge radius of $^{133}\mathrm{Sn}$ reveals a discontinuity in the odd-even staggering at $N=82$, similar to that observed at other major shell closures such as in Ca ($N=28$) and Pb ($N=126$)~\cite{Yan22}. The radius of $^{133}\mathrm{Sn}$ suggests that a kink occurs at $N=82$, which was previously identified for the even-even isotopes~\cite{Gorges2019}.

The VS-IMSRG calculations for $\Delta^{(3)}_{R}$ show improved agreement near these shell closures, although they tend to underestimate the staggering in the mid-shell where a greater collectivity is expected. The Fy(IVP) calculations accurately predict the inversion across $N=82$ but systematically overestimate the staggering throughout the entire chain due to a slightly too large proton pairing predicted in mid-shell tin isotopes. It is to be noted that the estimated uncertainty in the radii below $N=60$ in Fy(IVP) is greater than the staggering magnitude; therefore, the results for these isotopes are omitted in Fig.~\ref{Fig:Radii}(c). 

To gauge the impact of cross-shell excitation~\cite{Miyagi2020} across $N, Z=50$, we extended our  1.8/2.0(EM) VS-IMSRG calculations to a valence space based on the $^{80}$Zr core. 
This model space did, however, not include the $h_{11/2}$ orbit, which is known to be crucial for capturing the nuclear structure effects for isotopes with $N \gtrsim 64$~\cite{Qi2012}. Therefore, calculations with a $^{80}$Zr core were performed from $N=50$ up to $N=66$ only. The two choices of core should yield identical results if operators induced during the IMSRG flow were not truncated, so the discrepancy between them is indicative of the VS-IMSRG truncation error.
As shown in Fig.\,\ref{Fig:Radii}, the charge radii of the light tin isotopes predicted in VS-IMSRG show relatively low sensitivity to the assumed core. We therefore conclude the charge radii description can be captured by using a $^{100}$Sn core, with calculations performed in the  range from $N=50$ to $N=82$. (As we discuss below, this is not the case for $E2$ observables). 

\begin{figure}[h!]
\center
\includegraphics[width=0.97\columnwidth]{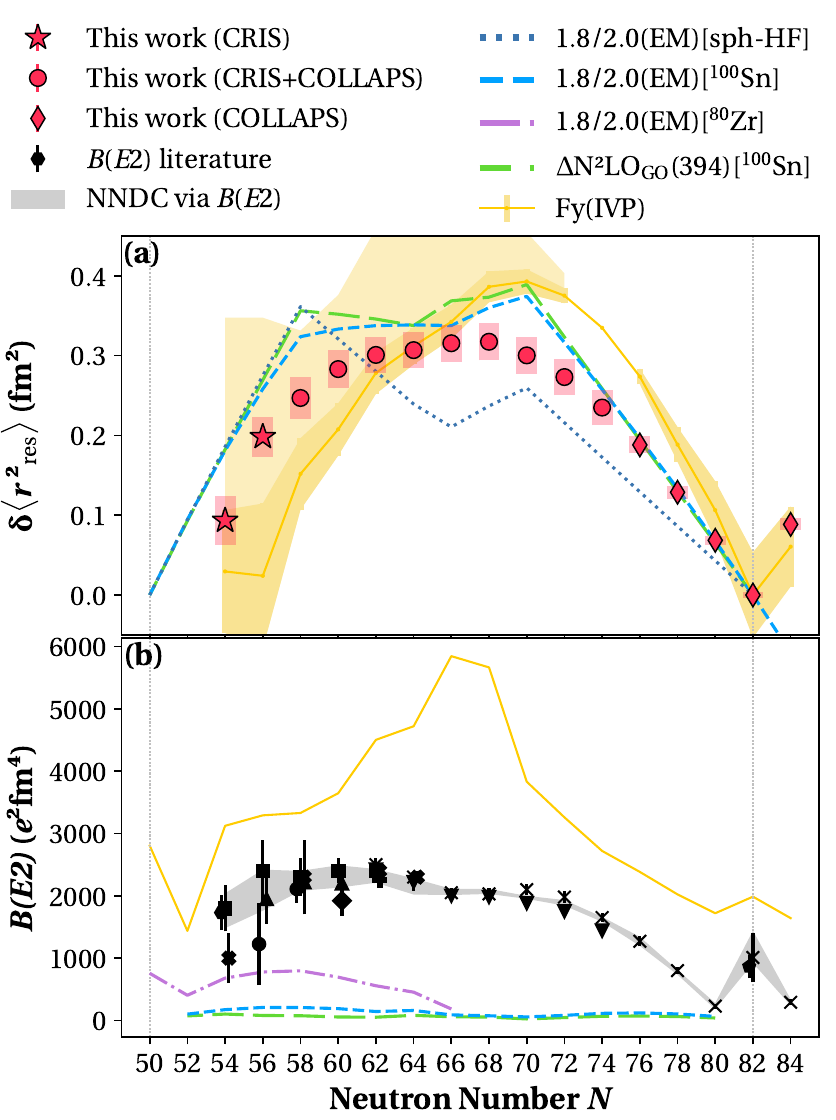}
\caption{(a) Residual differential mean-square charge radii $\delta \left\langle r_{\mathrm{res}}^2  \right\rangle $ obtained by the subtraction of a linear trend which intersects at the magic numbers $N=50, 82$ (see Ref.\,\cite{Karthein2024} for details). Experimental results are compared with DFT Fy(IVP) calculations, with systematic uncertainties shown as a yellow error band (darker shade represents the influence of ground-state correlations) and VS-IMSRG calculations using different Hamiltonians and different valence spaces (for details see text). The experimental and calculated $B(E2)$ values are shown in panel (b). Experimental $B(E2)$ values were taken from 
Refs.~\cite{Rosiak2018a,radford2003nuclear,guastalla2013,doornenbal2014,Ekstrom2008,allmond2011coulomb,varner2005coulomb,stelson1970static,back2013transition,siciliano2020pairing,BEENE2004471,RADFORD2005264}, shown with a 1-$\sigma$ error band from NNDC in Ref.\,\cite{pritychenko2016tables}. 
}
\label{Collectivity}
\end{figure}

We now focus on the even-even isotopes, which are simpler to interpret as the odd-neutron polarization is absent. The observed archlike trend in radii between shell closures is highlighted by removing the linear trend between the expected closed-shell isotopes $^{132}$Sn and $^{100}$Sn. The charge radius of $^{132}$Sn at the $N=82$ shell-closure is known. For $^{100}$Sn, the charge radius is estimated by fitting a parabolic function to the even-even radii down to $^{104}$Sn, following the same procedure described in Ref.\,\cite{Karthein2024}. In Fig. \ref{Collectivity}(a), the resulting archlike trend is shown after subtracting a bulk linear increase intersecting the charge radii at $N=50$ and $N=82$. 

This archlike behavior can be compared to the known data on the reduced transition probabilities $B(E2; 0^+_1 \rightarrow 2^+_1)$  in Fig.\,\ref{Collectivity}(b). These $B(E2)$ values provide insight into quadrupole collectivity along the isotopic chain. While the measured radii reveal a clearly reduced trend towards $N=50$, the trend for the $B(E2)$ values is less evident due to the large systematic uncertainty of the experimental values for the neutron-deficient isotopes. 
The $B(E2)$ values computed with Fy(IVP) were deduced from collective correlations, using the same approach as the correlation correction to the charge radii whereby the BCS approximation was used for simplicity \cite{Kluepfel_2008}. It is to be noted that these values here represent the sum over all $B(E2; 0^+ \rightarrow 2^+)$ transitions. Since the lowest $E2$ transition usually accounts for $\sim$90\% of the total $B(E2)$ strength, the Fy(IVP) values in Fig.\,\ref{Collectivity}(b) serve as upper estimates and should be reduced by 10\%. The Fy(IVP) results generally yield an archlike trend with a pronounced maximum at mid-shell and an only moderate reduction toward neutron-deficient isotopes. However, the mid-shell peak in $B(E2)$ is exaggerated in a similar fashion as the collective corrections for the radii. In addition, note that the DFT predictions of $B(E2)$ rates are still strongly varying among different functionals~\cite{Kluepfel_2008} and carry still rather large uncertainties, a feature which gives leeway for further improvement. A dedicated optimization to low-energy excitations is still a task for future development.

A similar archlike behavior of radii is obtained in the VS-IMSRG calculations, see  Fig.\,\ref{Collectivity}(a). To explore the origin of this trend, spherical Hartree-Fock calculations were performed in the equal filling approximation~\cite{Stro17}, in which all $m$-substates are filled equally, following the neutron orbitals in the order \( g_{7/2}, d_{5/2}, s_{1/2}, d_{3/2},\) and  \(h_{11/2} \). Hence, the observed trend is consistent with the charge radius following the neutron radius. The latter is mainly given by the sum of the single-particle mean-square radii weighted by their orbital occupation.
The effect of configuration mixing is thus visualized by the comparison with the spherical HF and full VS-IMSRG results. The spherical HF results are not sensitive to collective effects, yet these calculations predict an overall archlike trend in radii. We conclude that -- within the VS-IMSRG framework -- the evolution of charge radii between the shell closures is dominated by the uniform isotropic growth of the mean field seen by the protons and is not due to many-body correlations. 
This result emphasizes that care should be taken when studying the correlations between quadrupole collectivity and nuclear charge radii~\cite{sun2017correlating,caprio2022robust,otsuka2022moments,bro22}, as it is non-trivial to decouple the origin of different contributions to charge radii.

Finally, we note that the VS-IMSRG approach significantly underestimates the magnitude of the $B(E2)$ values, as seen in Fig. \ref{Collectivity}(b). As shown earlier~\cite{Hend18E2,Stro22E2} VS-IMSRG calculations based on spherical reference states typically underestimate absolute $B(E2)$ values due to missing particle-hole excitations which are needed to capture the full collective strength. The extended valence space using the $^{80}$Zr core shows an improvement, with a large increase for the magnitude of the predicted $B(E2)$ values with respect to the results using a $^{100}$Sn core, but still significantly underestimates the experimental values. This further highlights the importance of cross-shell excitations for describing the measured $B(E2)$ values of the neutron-deficient tin isotopes, as previously suggested by large-scale shell model calculations~\cite{togashi2018}. On the other hand, the charge radii are not affected by much as shown in panel (a) of Fig.~\ref{Fig:Radii}.
Evidently, the many-body dynamics driving the behavior of the charge radii in Fig.~\ref{Collectivity}(a) is distinct from that driving {$B(E2)$} in Fig.~\ref{Collectivity}(b). This emphasizes the bulk and collective nature of the charge radii and $B(E2)$ observables, respectively.

\textit{Conclusions}.-- We interpreted the laser spectroscopy measurements of charge radii for $^{104-134}$Sn by DFT and VS-IMSRG approaches. Both are able to describe the general behavior of the radii. While the general features are better predicted near the shell closures, improved theory is needed to reproduce the observed odd-even staggering in the mid-shell, where the significant quadrupole collectivity is expected. Both DFT and VS-IMSRG calculations predict appreciable contributions from beyond-mean-field correlations to the charge radii of the neutron-deficient tin isotopes.

As previously noted in Refs.~\cite{Parzuchowski2017,Hend18E2,Ver22,Stro22E2,LECHNER2023,Karthein2024}, reproducing collective effects such as quadrupole moments and $B(E2)$ values in VS-IMSRG remains a challenge. While an accurate reproduction of absolute charge radii in ab initio calculations is linked to developing improved nuclear forces with correct saturation, the underestimation of $B(E2)$ values is related to the lack of collective effects in the spherical VS-IMSRG employed in this work.

Within the VS-IMSRG framework, we have shown that the observed archlike trend of charge radii between shell closures is dominated by an expansion of the nucleus due to a gradual filling of neutron shells. This is in contrast with the conclusions obtained for $B(E2)$ values and other $E2$ observables~\cite{Karthein2024}, which require the inclusion of higher-order many-body correlations and/or non-spherical reference states. Anticipated charge radii and $B(E2)$ measurements reaching closer to $^{100}$Sn will be of great interest to complete our understanding of this frontier region of the nuclear chart. 

The data analysis used NumPy~\cite{numpy}, SciPy~\cite{scipy}, Pandas \cite{pandas1, pandas2},  iMinuit~\cite{iminuit1, iminuit2}., and qspec~\cite{Muller2025}. The figures were produced using Matplotlib~\cite{matplotlib}.

\begin{acknowledgments}
We thank the ISOLDE technical group for their
professional assistance. Useful comments from Ante Ravli\'c are gratefully 
acknowledged. This work was supported by ERC Consolidator Grant No.~648381 (FNPMLS); STFC grants ST/L005794/1, ST/L005786/1, (ST/L005670/1, ST/P004423/1, and Ernest Rutherford Grant No. ST/L002868/1; Projects GOA 15/010 and C14/22/104 from KU Leuven; the FWO-Vlaanderen project G080022N, G053221N, I002619N and I001323N (Belgium); the U.S. Department of Energy, Office of Science, Office of Nuclear Physics under grants DE-SC0021176 and DE-SC0013365, DE-SC0018083 and DE-SC0023175 (Office of Advanced Scientific Computing Research and Office of Nuclear Physics, Scientific Discovery through Advanced Computing); the Max-Planck Society; the Deutsche Forschungsgemeinschaft (DFG, German Research Foundation) -- Project-ID 279384907 -- SFB 1245; the European Research Council (ERC) under the European Union’s Horizon 2020 research and innovation programme (Grant Agreement No. 101020842); the German Federal Ministry for Education and Research under Contract No. 05P15RDCIA and 05P24RD4; the Helmholtz International Center for FAIR (HIC for FAIR) within the LOEWE program by the State of Hesse; the European Union Grant Agreement 654002 and 262010 (ENSAR); National Key R\&D Program of China (Contract No: 2018YFA0404403); the National Natural Science Foundation of China (No:11875073); NSERC under grants SAPIN-2018-00027, RGPAS-2018-522453, SAPIN-2024-0003, and the Arthur B. McDonald Canadian Astroparticle Physics Research Institute. TRIUMF receives funding via a contribution through the Digital Research
Alliance of Canada. Computations were performed with an allocation of computing resources on Cedar at WestGrid and Compute Canada. JK acknowledges support from a Feodor-Lynen postdoctoral research fellowship funded by the Alexander-von-Humboldt Foundation. We acknowledge the computing resources provided by MIT and the Texas A\&M high-performance computing cluster. TO, NS, and YT acknowledge the support by ``Program for Promoting Researches on the Supercomputer Fugaku'' (JPMXP1020200105, hp220174, hp210165). DTY acknowledges support from the Franco-Bulgarian Hubert Curien partnership Rila no. 51315QM and no. KP-06-RILA/4. TM and NS acknowledge the support by JST ERATO Grant No.~JPMJER2304, Japan. PGR thanks the regional computing center of the university Erlangen (RRZE) for support. This work was performed under the auspices of the U.S. Department of Energy, Office of Nuclear Physics (DOE NP) by Oak Ridge National Laboratory under Contract No. DE-AC05-00OR22725. DLB is supported in part through grants from the Romanian Ministry of Research, Innovation and Digitalization under contracts No. PN 23 21 01 06 and ELI-RO-RDI-2024-007 (ELITE). D. T. Y. acknowledges support from the Franco-Bulgarian Hubert Curien partnership Rila No. 51315QM
and No. KP-06-RILA/4. 
\end{acknowledgments}

\appendix
\section{Supplemental Material}


\subsection{VS-IMSRG calculations}
The VS-IMSRG calculations \cite{Hergert:2015awm,Stro19ARNPS} start from a Hamiltonian with two- and three-nucleon interactions represented in a space of 15 harmonic-oscillator shells with frequency $\hbar\omega=16$\,MeV and the sufficiently large truncation for 3N matrix elements $E_{\rm 3max}=24$~\cite{Miya22Heavy}. In the current study, we employed two chiral interactions, 1.8/2.0 (EM)~\cite{Entem2003,Hebe11fits} and $\Delta$N$^{2}$LO$_{\rm GO}$(394)~\cite{Jia20}. Combined with the ensemble normal ordering method~\cite{Stro17ENO} to capture effects of 3N forces between valence nucleons, we decouple a valence space Hamiltonian using the Magnus expansion detailed in Ref.\,\cite{Morris2015}, where operators are truncated at the two-body level, the IMSRG(2) approximation.
The radius operators are consistently transformed, and the charge radii are computed as in Ref.~\cite{Hagen2015}.

In the current study, three different valence spaces were adopted: The $sdg$ one-major shell space (above a $^{80}$Zr core), the neutron $\{2s_{1/2}, 1d_{3/2}, 1d_{5/2}, 0g_{7/2}, 0h_{11/2}\}$ orbits above a $^{100}$Sn core, and the neutron $\{2s_{1/2}, 1d_{3/2}, 1f_{7/2}, 0h_{11/2}\}$ above a $^{114}$Sn core.
The VS-IMSRG calculations were performed with the imsrg++ code~\cite{Stroberg++}, and the effective Hamiltonians for the $^{100}$Sn core space were diagonalized with KSHELL code~\cite{Shimizu2019}. Since the exact diagonalizations of the valence-space Hamiltonians within the $sdg$-shell space are not tractable, these calculations required the use of the recently developed quasi-particle vacua for nuclear shell-model calculations QVSM~\cite{Shimizu2021}.

\subsection{Experimental methodology and the determination of nuclear charge radii}

For the CRIS experiment, beams of stable and radioactive tin isotopes were produced from the bombardment of a heated LaC$_x$ target (1500-2000 $^{\circ}$C) with a 1.4 GeV proton beam at the ISOLDE-CERN facility. The nuclear reaction recoils escape the target material through diffusion and effusion and enter a heated transfer line to an ion source. There, the tin atoms were selectively ionized with the resonant ionization laser ion source (RILIS) \cite{fedosseev2017ion}, and electrostatically accelerated to 40 keV. The ion-beam is steered through the high-resolution mass separator (HRS) magnets for isotope separation ($\Delta m/m\sim6000$). The separated beam is then sent into a cooler buncher linear Paul trap (ISCOOL) \cite{mane2009ion}. The ion bunches are released with a bunch length of 5 $\mu s$, and transported to the CRIS beamline, described in detail elsewhere \cite{Kos20}. Subsequently the ions were neutralized with sodium vapour at 230-250\,$^{\circ}$C within a charge-exchange-cell (CEC). The fraction of neutral atoms within the beam varied between 10--30$\,$\%. Any remaining ions were removed by electrostatic deflectors after the CEC. The atoms entered the interaction region held at a high vacuum of $5\times10^{-8}$\,mbar to reduce the probability of collisional ionization of the neutral atom beam. Within the interaction region, each atom bunch was collinearly overlapped and synchronized in time with a series of laser pulses at 100\,Hz repetition rate, which step-wise excited and non-resonantly ionized the atomic bunch. The two atomic transitions probed for studying the isotope shifts are $5s^{2}5p^{2} \;  ^{3\!}P_{2} \rightarrow 5s^{2}5p6s\; ^{3\!}P_{2} $ (284\,nm) and $5s^{2}5p^{2} \;  ^{1\!}S_{0} \rightarrow 5s^{2}5p7s\; ^{1\!}P_{1}$ (281\,nm). The two transitions were studied using the three- and two-step ionization schemes, as described in Ref.\,\cite{Fredrik2020,parnefjord2021evolution}. These transitions were scanned with a speed below 5\,MHz/s across the resonance peak using an injection-seeded single-mode titanium sapphire laser, operating at a repetition rate of 1\,kHz, with a 20\,MHz bandwidth in the fundamental output. The resonantly ionized ions are deflected onto an ETP DM291 MagneTOF\textsuperscript{TM} detector with a timing resolution of 0.4\,ns for event-per-event ion counting.

\begin{figure}[]
\center
\includegraphics[width=1\columnwidth]{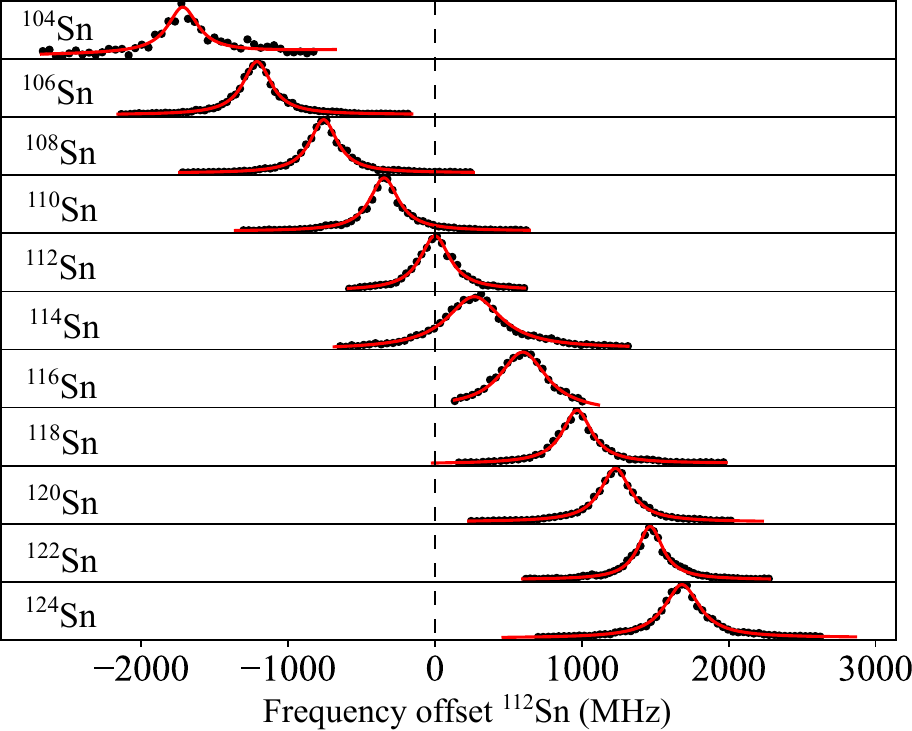}
\caption{Sample spectra of the even-even tin isotopes acquired with the $5s^{2}5p^{2} \; ^{3\!}P_{2} \rightarrow 5s^{2}5p6s \; ^{3\!}P_{2}$ (284\,nm) transition. The spectra are fitted using Lorentzian profiles with free linewidth as indicated with red lines.}
\label{Fig:spectrum}
\end{figure}

Example spectra of the Doppler-corrected resonances observed when probing the $5s^{2}5p^{2} \; ^{3}P_{2} \rightarrow 5s^{2}5p6s \; ^{3\!}P_{2}$ transition for the even-even ($I^{\pi}=0^+$) isotopes are presented in Fig.\,\ref{Fig:spectrum}, revealing the isotopic shift $\delta\nu^{A',A}$ which contains the information of the changes in the nuclear charge radii $\delta\left\langle r_\mathrm{c}^{2}\right\rangle^{A',A}$.

Throughout the experiment, the ion beam acceleration voltage was measured, allowing accurate Doppler-correction of the spectra. The wavelength of the scanned step was regularly measured using a WSU-2 HighFinesse wavemeter, which simultaneously recorded a reference wavelength from a stabilized laser diode (Toptica DLC DL PRO 780) locked to the Doppler-free hyperfine transition $5S_{1/2}(F=2)-5P_{3/2}(F'=3)$ of $^{87}$Rb (TEM CoSy) for correcting drifts of the wavemeter's interferometer. Furthermore, data taking on the isotopes of interest was interleaved throughout the experiment campaign with measurements of the selected reference isotope $^{112}$Sn to ensure any additional drifts were taken into account when determining the isotope shifts. 
\begin{NoHyper}
\begin{table*}[t]
\parbox{\textwidth}{\captionof{table}{Isotope shifts $\delta \nu_i^{\mathrm{ref},A}=\nu_i^A - \nu_i^{\mathrm{ref}}$ measured in the $5p^{2} \;  ^{3\!}P_{2} \rightarrow 5p6s\;^{3\!}P_{2}$ (284\,nm) and $5p^{2} \;  ^{1\!}S_{0} \rightarrow 5p7s \; ^{1\!}P_{1} $ (281\,nm) transitions at CRIS (ref=$^{112}$Sn) and the $5p^{2} \; ^{1\!}S_{0} \rightarrow 5p6s  \; ^{1\!}P_{1}$ (452.5\,nm) and $5p^{2} \; ^{3\!}P_{0} \rightarrow 5p6s \; ^{3\!}P_{1}$ (286.3\,nm) transitions at COLLAPS (ref=$^{124}$Sn). The statistical uncertainties are contained within parenthesis while the systematic uncertainties \rule{0mm}{3.5mm} are contained within square brackets. The changes in the mean-square charge radii $\delta\left\langle r_\mathrm{c}^2 \right\rangle^{124,A}_{\mu\mathrm{e}}$\rule{0mm}{3.5mm} obtained from muonic atom spectroscopy and elastic electron scattering are used for the King plots. For details see text.\rule{0mm}{3.5mm}}\label{table:IS_COLLAPS_CRIS}}
\begin{tabular*}{\textwidth}{@{\extracolsep{\fill}} c c@{} c@{} c@{} c@{}  l c @{} c @{}}
\hline \hline
     &  &          & \multicolumn{2}{c}{CRIS}                                                                &  & \multicolumn{2}{c}{COLLAPS}                                                                                      \\ \cline{4-5} \cline{7-8} 
$A$    & $I^{\pi}$ & {$\delta\left\langle r_\mathrm{c}^2 \right\rangle^{124,A}_{\mu\mathrm{e}}$} & \rule[-2mm]{0mm}{6mm} $\delta \nu^{112,A}_{\text{284\,nm}}$ (MHz) & $\delta \nu^{112,A}_{\text{281\,nm}}$ (MHz) &  & $\delta \nu^{124,A}_{\text{286\,nm}}$ (MHz) & $\delta \nu^{124,A}_{\text{453\,nm}}$ (MHz) \\ \hline \hline
104  & 0         & & $-1726.1\;(98)\;[55]$                            &                                            &  &                                                  \\
105  & \nicefrac{5}{2}$^+$ & & $-1523.5\;(61)\;[55]$                            &                                            &  &                                                  \\
106  & 0         & & $-1204.1\;(44)\;[55]$                            & $-902.4\;(77)\;[56]$                             &  &                                                        \\
107  & \nicefrac{5}{2}$^+$ & & $-1016.5\;(40)\;[55]$                            & $-783.9\;(40)\;[56]$                             &  &                                                       \\
108  & 0         & & $\;\;\;-752.2\;(40)\;[55]$                             & $-561.1\;(61)\;[56]$                             &  & $-2416.3\;(94)\;[98]$                                    \\
109  & \nicefrac{5}{2}$^+$   & & $\;\;\;-591.7\;(60)\;[55]$                             & $-449.9\;(74)\;[56]$                             &  & $-2238.6\;(19)\;[92]$                                    \\
110  & 0         & & $\;\;\;-348.4\;(34)\;[55]$                             & $-258.1\;(56)\;[56]$                             &  & $-2014.9\;(75)\;[84]$                                       \\
111  & \nicefrac{7}{2}$^+$   & & $\;\;\;-245.7\;(65)\;[55]$                             & $-177.3\;(42)\;[56]$                             &  & $-1910.6\;(36)\;[77]$                                     \\
112  & 0         & $-0.7480\;(77)$ & 0                                          & 0                                          &  & $-1652.1\;(56)\;[71]$                            & $-1383.3\;(22)\;[84]$                                            \\
113  & \nicefrac{1}{2}$^+$   & & $\;\;\;137.6\;(124)\;[55]$                             & $106.0\;(204)\;[56]$                             &  & $-1519.3\;(63)\;[65]$                            & $-1274.0(53)\;[37] $                                                  \\
114  & 0         & $-0.6016\;(77)$ & $\;\;\;314.4\;(\;\;48)\;[90]$                              &                                            &  & $-1335.6\;(62)\;[58]$                            & $-1115.3\;(19)\;[69]$                                                   \\
115  & \nicefrac{1}{2}$^+$   & & $\;\;\;417.2\;(\;\;22)\;[55]$                              &                                            &  & $-1250.4\;(28)\;[52]$                            & $-1044.2\;(12)\;[29]$                                                  \\
116  & 0         & $-0.4641\;(77)$ & $\;\;\;654.7\;(\;\;47)\;[90]$                              & $491.2\;(83)\;[56]$                              &  & $-1007.6\;(79)\;[45]$                            & $\;\;\;-841.8\;(14)\;[55]$                                                    \\
117  & \nicefrac{1}{2}$^+$   & & $\;\;\;754.8\;(\;\;43)\;[55]$                              &                                            &  & $\;\;\;-909.0\;(30)\;[40]$                               & $\;\;\;-759.3\;(17)\;[23]$                                                \\
118  & 0         & $-0.3278\;(77)$ &$\;\;\;957.8\;(\;\;30)\;[55]$                              & $754.2\;(59)\;[56]$                              &  & $\;\;\;-695.0\;(67)\;[33]$                               & $\;\;\;-585.6\;(19)\;[40]$                                                    \\
119  & \nicefrac{1}{2}$^+$   & & $1048.5\;(\;\;33)\;[90] $                            &                                            &  & $\;\;\;-620.4\;(25)\;[28]$                             & $\;\;\;-513.1\;(16)\;[16]$                                                     \\
120  & 0         & $-0.2022\;(80)$ & $1225.3\;(\;\;22)\;[55]$                             &                                            &  & $\;\;\;-448.6\;(80)\;[22]$                             & $\;\;\;-360.5\;(13)\;[27]$                                                     \\
121  & \nicefrac{3}{2}$^+$   & & $1302.6\;(332)\;[55]$                            &                                            &  & $\;\;\;-355.1\;(37)\;[17]$                             & $\;\;\;-289.9\;(40)\;[09]$                                                 \\
122  & 0         & $-0.0930\;(77)$ & $1459.8\;(\;\;32)\;[55]$                             & $1126.4\;(48)\;[56]$                             &  & $\;\;\;-206.3\;(75)\;[10]$                             & $\;\;\;-167.3\;(16)\;[13]$                                                    \\
123  & \nicefrac{11}{2}$^-$  & & $1533.4\;(153)\;[55]$                            & $1178.7\;(52)\;[56]$                             &  & $\;\;\;-135.5\;(31)\;[05]$                             & $\;\;\;-114.5\;(23)\;[03]$                                                     \\
124  & $0 $        & $0.0\;(0)$ & $1678.3\;(\;\;34)\;[55]$                             & $1276.5\;(52)\;[56]$                             &  & $0$                                          & $0$                                                   \\
125  & \nicefrac{11}{2}$^-$  & &                                           &                                            &  & $\;\;\;\;\;\;61.0\;(24)\;[05]$                               & $\;\;\;48.4\;(12)\;[03]$                                                       \\
126  & 0         & &                                           &                                            &  & $\;\;\;189.4\;(47)\;[12]$                              & $147.2\;(13)\;[13]$                                                       \\
127  & \nicefrac{11}{2}$^-$  &  &                                          &                                            &  & $\;\;\;246.2\;(32)\;[16]$                              & $190.5\;(26)\;[09]$                                                    \\
128  & 0         &  &                                         &                                            &  & $\;\;\;337.3\;(49)\;[21]$                              & $274.5\;(31)\;[12]$                                                     \\
129  & \nicefrac{3}{2}$^+$   &   &                                         &                                            &  & $\;\;\;350.2\;(47)\;[26]$                              & $268.2\;(41)\;[15]$                                                    \\
130  & $0$         &  &                                          &                                            &  & $\;\;\;491.2\;(30)\;[31]$                              & $395.0\;(18)\;[18]$                                                     \\
131  & \nicefrac{3}{2}$^+$   &     &                                       &                                            &  & $\;\;\;466.7\;(31)\;[36]$                              & $353.0\;(22)\;[20]$                                                      \\
132  & $0$         &    &                                        &                                            &  & $\;\;\;624.0\;(32)\;[42]$                              & $497.7\;(11)\;[50]$                                                     \\
133  & \nicefrac{7}{2}$^-$   &     &                                       &                                            &  & $\;\;\;868.5\;(29)\;[46]$                              & $705.8\;(16)\;[26]$                                                      \\
134  & 0         &    &                                        &                                            &  & $1196.5\;(74)\;[51]$                             & $981.0\;(36)\;[62]$                                                      \\ \hline \hline
\end{tabular*}
\end{table*}
\end{NoHyper}
\twocolumngrid

The $^{3\!}P_{2} \rightarrow {}^{3\!}P_{2}$ (284\,nm) transition in tin was predominantly used throughout the experiment, allowing spectroscopy of 21 tin isotopes ranging from $^{104}$Sn to $^{124}$Sn. Complementary measurements were performed using the $^{1\!}S _{0} \rightarrow {}^{1\!}P_{1}$ (281\,nm) transition down to $^{106}$Sn. During the spectroscopy of $^{114,116}$Sn using the $^{3\!}P_{2} \rightarrow {}^{3\!}P_{2}$ transition, the laser power of the first step was increased by a factor five for enhancing ionization efficiency while increasing the linewidth from 230\,MHz to 350\,MHz 
FWHM. Reference measurements of both $^{112}$Sn and $^{120}$Sn were performed under these identical conditions. Resonances were fitted using a Lorentz profile with free linewidth, since no Gaussian contribution was observed, with an average linewidth of 230\,MHz for all isotopes except those acquired using a higher laser power. As an example, the fitted spectra extracted using the $^{3\!}P _{2} \rightarrow {}^{3\!}P_{2}$ transition of the even-even isotopes are presented in Fig.\,\ref{Fig:spectrum}.

The determined isotope shifts relative to $^{124}$Sn are provided in Table~\ref{table:IS_COLLAPS_CRIS} for both transitions. In order to obtain $\delta \!\left\langle r_\mathrm{c}^2 \right\rangle^{124,A}$ from the isotope shifts measured at CRIS relative to $^{112}$Sn, the following procedure was used. First, the isotope shifts relative to $^{124}$Sn were calculated and the statistical uncertainties of the isotope of interest and that of the new reference $^{124}$Sn were added in quadrature. To determine the total uncertainty, we combined the statistical uncertainty with the systematic uncertainty specific to each isotope. In this particular case, the systematic uncertainty was not derived from a potential voltage offset. Instead, it stemmed from laser drifts that could occur during the measurement of different isotopes, whether in positive or negative directions. For the isotope $^{112}$Sn, we used the systematic uncertainty associated with $^{124}$Sn. This total uncertainty was then used to perform the King plot for the individual transitions. For the COLLAPS data, only the statistical uncertainty was taken into account, since the systematic uncertainty is dominated by the scaling uncertainty of the voltage divider that transfers into a mass-shift like behavior in the isotopic chain and is, thus, captured by the King-plot procedure. 

\begin{table}[]
\caption{King-fit parameters in the $5p^{2} \;  ^{3\!}P_{2} \rightarrow 5p6s\,^{3\!}P_{2}$ (284\,nm), the $5p^{2} \;  ^{1\!}S_{0} \rightarrow 5p7s \; ^{1\!}P_{1} $ (281\,nm),  the $5p^{2} \; ^{1\!}S_{0} \rightarrow 5p6s  \; ^{1\!}P_{1} $ (452.5\,nm) and the $5p^{2} \; ^{3\!}P_{0} \rightarrow 5p6s \; ^{3\!}P_{1}$ (286.3\,nm) transitions. $M$ is the physically relevant mass shift constant, and $M_\alpha$ is the mass shift parameter for the corresponding value $\alpha$, which has been optimized to remove the correlation between the slope and the offset term \cite{Gorges2019}.}
\begin{tabular}{ccccc}
\hline \hline
Transition & 284 nm & 281 nm & 286 nm & 453 nm \\ \hline
$F$ (GHz/fm$^2$) & 2.51 (53) & 2.68 (72) & 2.42 (47) & 2.74 (57) \\
$M$ (GHz u) & -228 (447) & -838 (610) & -492 (397) & -456 (480) \\
$M_\alpha$ (GHz u) & 1888 (17) & 1428 (21) & 1543 (15) & 1850 (17) \\
$\alpha$ ( fm$^2$ u) & 843 & 846 & 841 & 842  \\ \hline \hline
\label{table:atomic_factors}
\end{tabular}
\end{table}

The charge radii of the stable isotopes for the King plots are obtained based on the data provided in the book of Fricke and Heilig \cite{50-Sn}: The Barrett radii determined from muonic atom spectroscopy were divided by the isotope-dependent $V^\mathrm{e}_2$ factors ($V_2^{\mathrm{e}}=R_{k \alpha}^{\mathrm{e}} /\left\langle r_\mathrm{c}^{\mathrm{2}}\right\rangle_{\mathrm{e}}^{1/\mathrm{2}}$) obtained from elastic electron scattering to obtain the mean-square charge radii. From these, the $\delta\left\langle r_\mathrm{c}^2 \right\rangle_{\mu\mathrm{e}}$ of the stable even isotopes were obtained. We note that two sets of Barrett radii are listed in \cite{50-Sn}, which have been treated identically and the changes in the mean-square charge radii were finally averaged to arrive at a single set. These are listed in Table~\ref{table:IS_COLLAPS_CRIS} and used in the four King plots.

The final value for $\delta \!\left\langle r_\mathrm{c}^2 \right\rangle^{124, A}$ as listed in the main manuscript in Table I is then obtained as the weighted mean of the King-plot results of the four transitions. The final uncertainty has been chosen as the smallest uncertainty that was obtained from one specific transition, but no further reduction was considered due to the strong correlations via the respective field shift uncertainty. It was found that the size of this uncertainty covers approximately the observed range of variation of the results in the four transitions.

The mass shift and field shift factors obtained in the King plots, which are required to obtain the $\delta \!\left\langle r_\mathrm{c}^2 \right\rangle^{124, A}$ values in the respective transitions, are summarized in Table \ref{table:atomic_factors}. The differential mean-square charge radii and their uncertainties for each transition individually can be obtained using the values in Table~\ref{table:IS_COLLAPS_CRIS} and Table~\ref{table:atomic_factors}, according to $\delta\left\langle r_\mathrm{c}^2\right\rangle^{124,A}=(\delta \nu^{124,A}-K_\alpha \mu) / F+\alpha \cdot \mu$ with the mass factor $
\mu=\left(m_{A}-m_{124}\right) /\left(m_A \times m_{124}\right)$. In the evaluation of the uncertainty on $\delta \!\left\langle r_\mathrm{c}^2 \right\rangle^{124, A}$, the optimized $M_\alpha$ ensures that no correlations between $F$ and $M_\alpha$  exist in the error propagation. \\

\FloatBarrier
\bibliography{Bibliography}

\end{document}